\documentclass[prb,twocolumn,aps]{revtex4}
\usepackage{epsfig,epsf,psfrag}
\usepackage{graphicx}
\usepackage{epic,eepic}
\usepackage{color,pstcol}
\begin{document}
\title{$\pi$ junction qubit in monolayer graphene}
\author{Colin Benjamin and Jiannis K. Pachos}
\affiliation {Quantum Information Group, School of Physics and
Astronomy, University of Leeds, Woodhouse Lane, Leeds LS2 9JT,
UK.}
\begin{abstract}
  We propose to combine the advantages of graphene, such as easy
  tunability and long coherence times, with Josephson physics to
  manufacture qubits.  If these qubits are built around a $0$ and
  $\pi$ junction they can be controlled by an external flux.
  Alternatively, a d-wave Josephson junction can itself be tuned via a
  gate voltage to create superpositions between macroscopically
  degenerate states. We show that ferromagnets are not required for
  realizing $\pi$ junction in graphene, thus considerably simplifying
  its physical implementation.  We demonstrate how one qubit gates,
  such as arbitrary phase rotations and the exchange gate, can be
  implemented.
\end{abstract}
\maketitle
\section{Introduction}
Graphene, a monatomic layer of graphite exhibits promising electronic
properties that can be employed for quantum technologies\cite{castro}.
Characteristically, its low energy excitations are described by the
Dirac equation, it has a zero band gap, electronic speeds can reach a
hundredth of the speed of light and it supports long range phase
coherence. However it has not yet been utilised to create qubits
suitable for quantum computation, apart from a proposal which meshes
it with bilayer structures\cite{loss}. Here we show that a key
ingredient of Josephson qubits, a $\pi-$junction\cite{golubov} can be
easily generated in graphene by application of a gate voltage alone.
We establish a parametric regime for observing this effect and show
how to manufacture qubits. These Josephson qubits can be used to
perform single quantum gates, such as the phase and exchange gates.
This opens up the possibility of employing graphene and utilizing its
advantages for quantum information processing\cite{nori}.
\begin{figure}
  \centerline{\includegraphics[width=8cm,height=4.0cm]{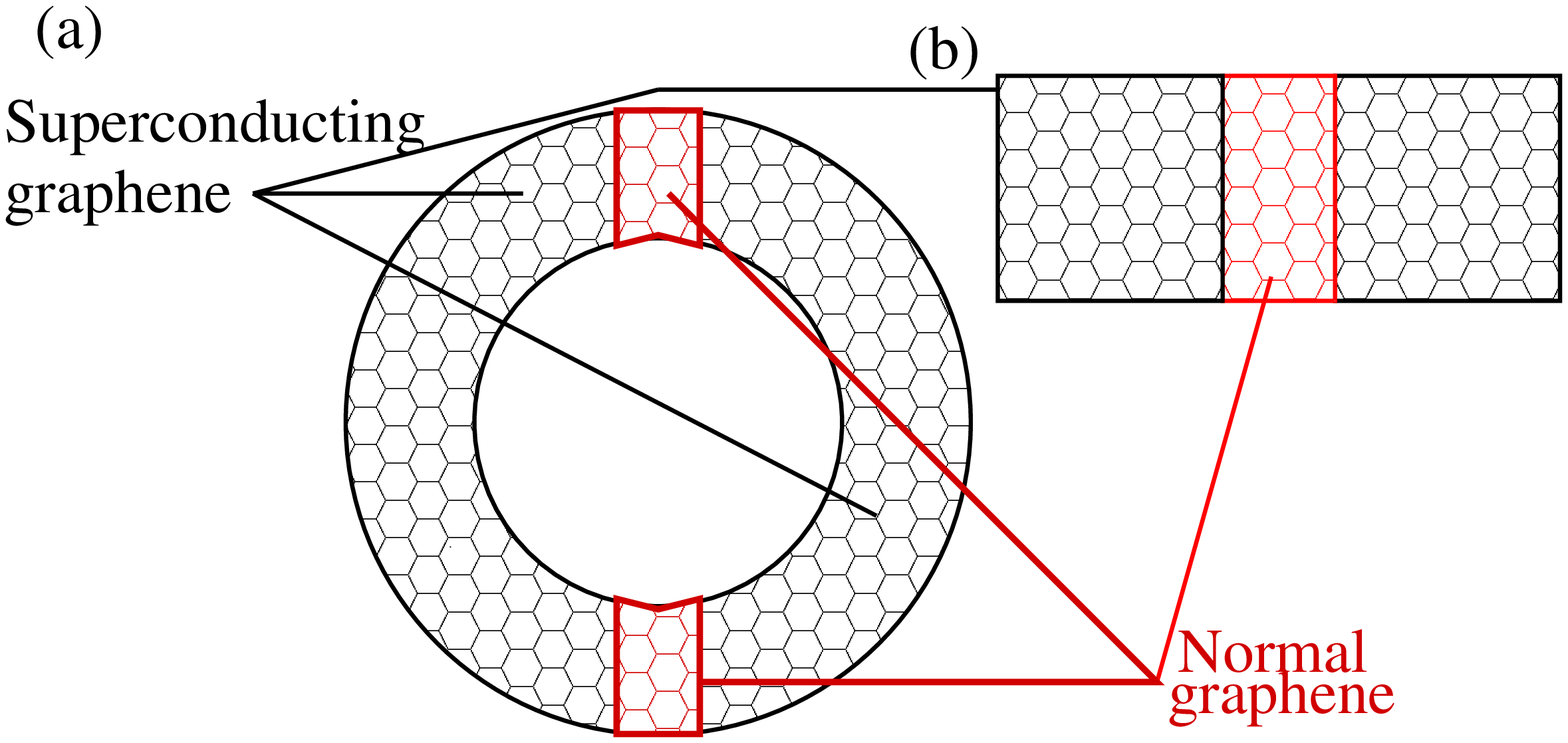}}
  \caption{(Color online) An overview of the set-up. (a) Two semicircular d-wave
    superconducting graphene strips (Gs) with normal graphene layers
    on top and bottom enclosing a magnetic flux, $\Phi$. By the
    application of suitable gate voltages to the normal graphene strip
    the junctions are tuned to either $\phi=0$ or $\pi$ phase shift.
    (b) A graphene d-wave Josephson junction. For relatively small intervening length   between the superconducting graphene one can
    have situations wherein degenerate ground states are formed and
    are pliable to external control via a gate voltage.
    \label{scheme}}
\end{figure}

The physical system we employ consists of a graphene substrate with
superconducting correlations induced in sections via the proximity
effect~\cite{gueron} or by turning graphene
superconducting\cite{carlson} via doping. It comprises of two d-wave
Josephson junctions (distinguished by their ground states, one at a
phase difference $\phi=0$ and the other at $\phi=\pi$), arranged as in
Fig. 1(a). The total energy of the system is controlled by the flux,
$\Phi$, that passes through the ring. The reversal of super-current in
a Josephson device, where the free energy has global minima at phase
difference $\phi=\pi$, is referred to as $\pi$ shift. The
corresponding Josephson junction is termed a $\pi$ junction. This is
in contrast to a $0$ junction wherein the free energy has a global
minimum at phase difference $\phi=0$~\cite{benjamin}. To be able to
encode a qubit we have to construct a $\pi$ junction and integrate it
with the rest of our device (the $0$ junction). A $\pi$ junction is
needed to create a doubly degenerate ground state, where a qubit is
encoded. Here we demonstrate that a $\pi$ junction can be identified
in our system without the need of any ferromagnetic
elements\cite{linder}, thus greatly simplifying its experimental
implementation. In Fig.  1(b) we depict a simple d-wave graphene
Josephson junction, which has two degenerate ground states, that can
encode a qubit. In particular, we prove that a complete set of single
qubit gates can be efficiently implemented demonstrating that our
proposal is promising for quantum computation.

\begin{figure}
  \centerline{\includegraphics[width=8cm,height=6cm]{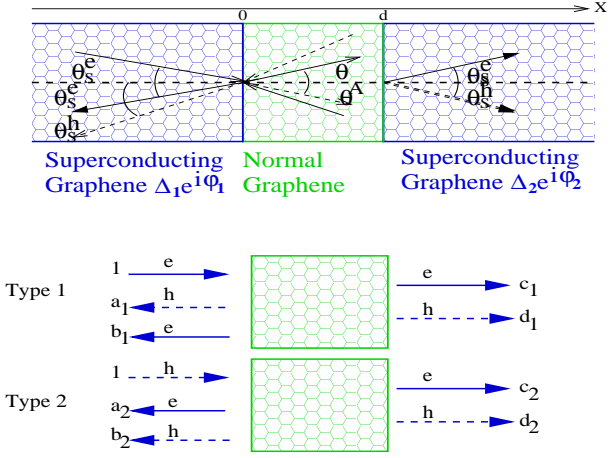}}
  \caption{(Color online) The Furusaki-Tsukada approach and the processes involved.
    Top: $\theta^{e}_{S}$ is the angle of incidence of electron-like
    quasiparticle, while $-\theta^{e}_{S}$ is the angle of its
    reflection. Hole-like quasiparticle are Andreev reflected at angle
    $\theta^{h}_{S}$. In the normal region electron and holes are
    transmitted and incident with angles $\theta$ and $\theta^A$.
    Bottom: In type 1 process an electron-like quasiparticle is
    incident from the left, while in type 2 process a hole-like
    quasiparticle in incident from the left. $a_1$, $b_2$, $d_1$ and
    $d_2$ are amplitudes of hole-like quasiparticle, while $a_2$,
    $b_1$, $c_1$ and $c_2$ are scattering amplitudes for electron-like
    quasi-particles. \label{type}}
\end{figure}
In Fig. \ref{type}, we show our graphene $\pi$ junction set-up.
  It is known that with s-wave
superconductors a $\pi$ junction is not possible\cite{been-tit}.
However, a Josephson junction with d-wave superconductors can
exhibit a $\pi$ shift\cite{sigrist}. Thus, we
consider d-wave correlations in the superconducting segments (see Fig. 1).

\section{Theory}
The kinematics of quasi-particles in graphene is
described by the Dirac-Bogoliubov-de Gennes
equation\cite{graphene-rmp}, which assumes the form
\begin{equation}
  \left( \begin{array}{cc}
    \hat{H} - E_F \hat{I} & \Delta \hat{I}\\
    \Delta^{\dagger} \hat{I} & E_F \hat{I} - \hat{T} \hat{H} \hat{T}^{- 1}
  \end{array} \right) \Psi = E \Psi,
\end{equation}
where $E$ is the excitation energy, $\Delta$ is the
superconducting gap of a d-wave superconductor, $\Psi$ is the
wavefunction and $\hat{\cdot}$ represents $4 \times 4$ matrices.
In the above equation
\begin{equation}
  \hat{H} = \left( \begin{array}{cc}
    H_+ & 0\\
    0 & H_-
  \end{array} \right), \hspace{0.25em} \hspace{0.2em} H_{\pm}
  = - i \hbar v_F (\sigma_x \partial_x \pm \sigma_y \partial_y) + U.
\end{equation}
Here $\hbar, v_F$ (set equal to unity hence forth) are the Planck's
constant and the energy independent Fermi velocity for graphene, while
the $\sigma$'s denote Pauli matrices that operate on the sub-lattices
$A$ and $B$.  The electrostatic potential $U$ can be adjusted
independently via a gate voltage or doping. We assume $U = 0$, in the
normal region, while \ $U = - U_0$ in the superconducting graphene. In
our work we consider $U_{0}=100\Delta$. Further we choose d-wave
superconducting correlations which imply a type II (or high $T_c$)
superconductor. This is most likely to be observed in
graphene\cite{carlson}. The subscripts of Hamiltonian $\pm$ refer to
the Fermi points $K_+$ and $K_-$ in the Brillouin zone. $T = - \tau_y
\otimes \sigma_y C,$ ($C$ being complex conjugation) is the time
reversal operator, with $\tau$ being Pauli matrices that operate on
the $\pm$ space and $ \hat{I}$ is the identity matrix.

To calculate the Josephson supercurrent, Free energy and show the
formation of a $\pi$ junction we proceed by first calculating the
scattering wave functions of our system. Let us consider (Type 1
scenario in Fig. 2)an incident electron-like
quasiparticle\cite{graphene-sudbo} from the left superconductor with
pairing gap $\Delta(\theta^{+}) e^{i \phi^{+}_1}$ ($x < 0$) and energy
$E$. For a right moving electron-like quasiparticle with an incident
angle $\theta$ the eigenvector and corresponding momentum read
$\Psi^e_{S_{1} +}=[u_{e}, u_{e}e^{i \theta^+},v_{e}e^{-i\phi^{+}_{1}},
v_{e}e^{i( \theta^+-\phi^{+}_{1})}]^T e^{iq^e \cos \theta^+ x}, q^e
=(E_F + U_0 +\sqrt{E^2 - |\Delta(\theta^{+})|^{2}})$. A left moving
electron-like quasiparticle is described by the substitution $\theta
\rightarrow \pi - \theta$. If Andreev-reflection takes place, a left
moving hole-like quasiparticle is generated with energy $E$, angle of
reflection $\theta_{}^-$ and its corresponding wavefunction is given
by $\Psi^h_{S_{1} -}=[v_{h},-v_{h}e^{- i \theta^-},
u_{h}e^{-i\phi^{-}_{1}} , -u_{h}e^{- i(\theta^{-}+\phi^{-}_{1})}]^T
e^{- iq^h \cos \theta^- x},q^h = (E_F + U_0 - \sqrt{E^2 -
  |\Delta(\theta^{-})|^{2}})$.  The quasi-particle wave-vectors can
also be expressed as $q^{e/h}=E_{F}+U_{0}\pm 1/\xi$, where $\xi$ is
the coherence length. For the Dirac-Bogoliubov de Gennes equations to
hold the Fermi wavelength in the superconductor $1/(E_{F}+U_{0})$
should be much smaller than the coherence length. The superscript e
(h) denotes an electron-like (hole-like) excitation. Since
translational invariance in the $y$-direction holds the corresponding
component of momentum is conserved. This condition allows for the
determination of the Andreev reflection angle $\theta^-$ through $q^h
\sin (\theta_{}^-) = q^e \sin (\theta^+)$.  The coherence factors are
given by $u_{e/h} = \sqrt{(1 + \sqrt{1 - |\Delta(\theta^{\pm})|^{2} /
    E^2}) / 2}$, $v_{e/h} = \sqrt{(1 - \sqrt{1 -
    |\Delta(\theta^{\pm})|^{2} / E^2}) / 2}$. We have also defined
$\theta^+ = \theta^e_S,$ $\theta^- = \pi - \theta^h_S$, where the
angles are defined in Fig.~\ref{type}. In our study we have d-wave
superconductors, thus
$\Delta(\theta^\pm)=\Delta\cos(2\theta^{\pm}-2\gamma)$ and the
macroscopic phase is
$e^{i\phi^{\pm}_{1/2}}=e^{i\phi_{1/2}}\frac{\Delta(\theta^\pm)}{|\Delta(\theta^\pm)|}$.
We choose the superconductor oriented along the $110$ direction,
implying $\gamma=\pi/4$.

In the normal region the eigenvector and corresponding momentum of a
right moving electron with an incident angle $\theta$ read: $ \psi^e_+
= [1, e^{i \theta}, 0, 0]^T e^{ip^e \cos \theta x}, \hspace{0.25em}
\hspace{0.25em} \hspace{0.25em} p^e = (E + E_F).$ A left moving
electron is described by the substitution $\theta \rightarrow \pi -
\theta$. If Andreev-reflection takes place, a left moving hole is
generated with energy $E$, angle of reflection $\theta_A$ and its
corresponding wave function is given by- $\psi^h_- = [0, 0, 1, e^{- i
  \theta_A}]^T e^{- ip^h \cos \theta_A x}, \hspace{0.25em}
\hspace{0.25em} \hspace{0.25em} p^h = (E - E_F).$ The transmission
angles $\theta$ and $\theta_A$ for the electron-like and hole-like
quasi-particles are given by $q^{e} \sin \theta^{e}_S = p^e \sin
\theta$ and $q^{e} \sin \theta^{e}_S = p^h \sin \theta_A$.

The full wave function in the type 1 scenario can be written as
below for the various regions
\begin{eqnarray}
  \psi_{S_{1}}&=&\Psi^{e}_{S_{1}+}+b_{1}\Psi^{e}_{S_{1}-}+a_{1}\Psi^{h}_{S_{1}-},
  \,\, x<0,\nonumber\\
  \psi_{N}&=&p\psi^{e}_{+}+q\psi^{e}_{-}+m\psi^{h}_{+}+n\psi^{h}_{-},\,\,0<x<d,\nonumber\\
  \psi_{S_{2}}&=&c_{1}\Psi^{e}_{S_{2}+}+d_{1}\Psi^{h}_{S_{2}+}, \,\, x>d.
\end{eqnarray}
Matching the wave functions at the interfaces one can solve for the
amplitudes of reflection $a_{1}$, $b_{1}$, $c_{1}$ and $d_{1}$.
Similarly, one can write the wave functions in case of type 2 scenario
(hole incident from the right) and calculate the amplitudes $a_{2}$,
$b_{2}$, $c_{2},$ and $d_{2}$.  The detailed balance for the
amplitudes are verified as follows
\begin{eqnarray}
C a_1(\phi,E)&=&C' a_2(-\phi,E),\nonumber\\
b_{i}(\phi,E)&=&b_{i}(-\phi,E) (i=1,2),
\end{eqnarray}
with $C=\frac{\Omega_{n,-}}{|\Delta(\theta^{-})|}\cos\theta^h_S$,
and $C'=\frac{\Omega_{n,+}}{|\Delta(\theta^{+})|}\cos\theta^e_S$.
Following the procedure established in Ref.\cite{fur-tsu} and
employing analytic continuation $E \rightarrow iw_n$ the dc
Josephson current is calculated as
\begin{eqnarray}
  I(\phi)&=&\,\,\,\,\sum_{w_n}\frac{e}{2\beta\hbar}\int^{\pi/2}_{-\pi/2}
  \,\,\,\,[\frac{a_{1}(\theta^{+},\phi,iw_n)}{C'}\,\nonumber\\
  & &  \hspace{0.25em} \hspace{0.25em} -\,\frac{a_{2}(\theta^{+},\phi,iw_n)}{C}]\cos(\theta^e_S)d\theta^e_S,\nonumber\\
  &=&\,\,\,\,\sum_{w_n}\frac{e}{2\beta\hbar}\,\,\,\,\int^{\pi/2}_{-\pi/2}
  \frac {|\Delta(\theta^+)|}{\Omega_{n,+}}[{a_{1}(\theta^{+},\phi,iw_n)}\,\nonumber\\
  & &\hspace{0.25em} \hspace{0.25em} -\,{a_{1}(\theta^{+},-\phi,iw_n)}]d\theta^e_S.
  \label{eq:Ij}
\end{eqnarray}
where $\beta=1/k_{B}T,
\Omega_{n,\pm}=\sqrt{w^{2}_{n}+|\Delta(\theta^{\pm})|^2}$ and
$w_{n}=\pi k_{B} T (2n+1)$, $n=0, \pm 1, \pm 2, ...$.

The above equation has a simple physical
interpretation~\cite{fur-tsu}. Andreev reflection is equivalent to the
breaking up or creation of a Cooper pair. The scattering amplitude
$a_1$ describes the process in which an electron-like quasiparticle
coming from the left superconducting graphene strip ($x<0$) is
reflected as a hole-like quasiparticle.  The amplitude $a_2$
corresponds to the reverse process in which a hole-like quasiparticle
is reflected as an electron-like quasiparticle. This implies that
$a_1$ and $a_2$ correspond to the passage of a Cooper pair to the left
and right respectively, hence, the dc Josephson current is
proportional to $a_{1}-a_{2}$.  Further, the dc Josephson current is
an odd function of the phase difference, $\phi$, as seen by the
detailed balance condition,
$a_{2}(\phi,iw_{n})/C=a_{1}(-\phi,iw_{n})/C'$. To calculate the
Josephson current one thus takes the difference between the amplitudes
$a_1$ and $a_2$ and then sums over the energies. In this approach we
account for all the energies both bound states and the continuum. 
Eq.\ref{eq:Ij} can be simplified as-
\begin{eqnarray}
  I(\phi)&=&\,\,\,\,\sum_{w_n}\frac{e}{2\beta\hbar}\,\,\,\,\int^{\pi/2}_{-\pi/2}
  \frac {|\Delta(\theta^+)|}{\Omega_{n,+}}[2iJ]d\theta^e_S, \mbox{ and }\nonumber\\
J&=&\frac{A\sin(\phi)+B\sin(2\phi)}{A'+2B'\cos(\phi)+2C'\cos(2\phi)}
\label{eq:Ij-simp}
\end{eqnarray}
In Eq.\ref{eq:Ij-simp}, $A, B, A', B',$ and $ C'$ are functions of $\theta^{+}, iw_n, E_f \mbox {and } d$. 
The Free energy of the Josephson junction can then be calculated as
\begin{equation}
  F(\phi)=\frac{1}{2\pi}\int_{0}^{\phi} I(\phi') d\phi'.
  \label{eq:Free}
\end{equation}
\section{$\pi$-junction}
\begin{figure}
  \centerline{\includegraphics[width=9cm,height=8cm]{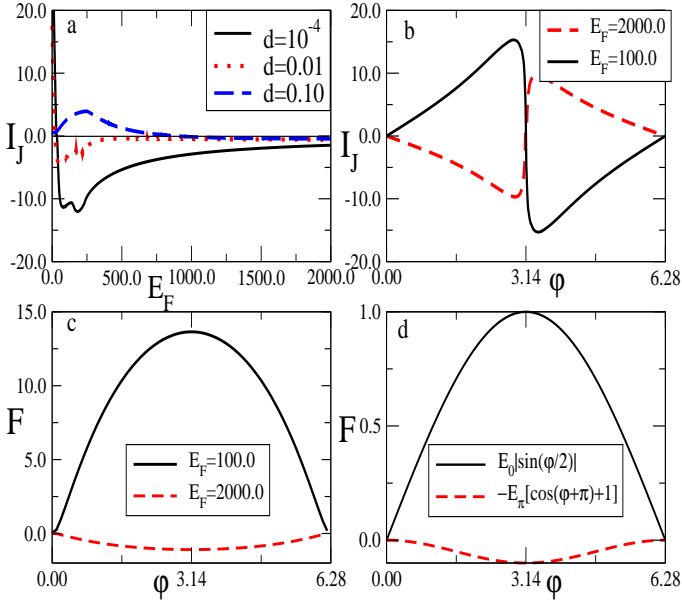}}
  \vskip 0.2 in\caption{(Color online) (a) Current (in units of $e\Delta/\hbar$ and
    normalized by $1/\beta$ throughout in all succeeding figures)
    versus Fermi energy, $E_F$, at phase difference $\phi=\pi/2$, for
    different values of width $d$(in units of $ \hbar v_{F}/\Delta$),
    $U_{0}=100 \& k_{B}T=0.0001$ in this and all succeeding figures.
    (b) Current versus phase, where the length of normal graphene
    strip is $d=0.1$, the dashed (RED) line is multiplied by a factor
    of ten for better visibility.(c) Free energy (normalized by
    $1/\beta$) of $G_{S}-G_{N}-G_{S}$ junction versus phase difference
    for different Fermi energies with 0 junction ($E_F=2000$ red
    dashed line) and $\pi$ junction ($E_F=100$ black solid line) and
    length of normal graphene strip $d=0.1$. (d) The approximate forms
    for the $0$ and $\pi$ junction energies are in good agreement with
    the real free energies and are used in analyzing the graphene
    Josephson qubit.\label{J-d}}
\end{figure}

Now we illustrate the results for the Josephson current as function of
the length of the normal graphene interlude as well as the phase
difference across the two superconducting graphene strips. The
calculations are performed by treating Eqs.~(\ref{eq:Ij})
and~(\ref{eq:Free}) numerically and the derived results hold for the
$T\rightarrow 0$ temperature limit.  Fig.~\ref{J-d}(a) shows the
Josephson current as function of the Fermi energy, in the normal
graphene strip, for different lengths of the normal graphene layer.
Note that Fermi energy is easily controllable in graphene. The plot
shows that for extremely small length of normal graphene layer the
Josephson current is negative for a wide range of Fermi energy,
implying a $\pi$ shift, while for larger intervening normal layers the
Josephson current changes sign at larger values of the Fermi energy.
One important fact to note is that for increased $d$ the current
decreases, which is in agreement with past Josephson works. Another
observation from Fig.~\ref{J-d}(a) is that at large Fermi energy the
Josephson supercurrent becomes independent of $E_f$. The explanation
for this is- when $E_F >> E, \Delta$, the angles for electron and
hole-like quasi-particles are
$\theta_{S}^{e}=\theta_{S}^{h}=\theta=-\theta_A$. With this condition,
the factor $J$ from Eq.\ref{eq:Ij-simp}, the Josephson supercurrent shorn of all prefactors, reduces to-

\begin{equation}
J=\frac{-ie^{-i\gamma}\sin(2\theta)}{E(h^{2}+e^{-2i\gamma}g^{2})}
\label{eq:J}
\end{equation}

In the above equation,
$\gamma=(p_{e}+p_{h}) d \cos(\theta)=E d \cos(\theta), h=(E-x)/2E,
g=(E+x)/2E, x=\sqrt{E^{2}-\sin(\theta)^{2}}$.  Thus in this limit the
Josephson supercurrent becomes completely independent of $E_F$.
Further, for $ d \rightarrow 0$ one can clearly see from
Fig.~\ref{J-d}(a) that the Josephson supercurrent becomes completely
negative, this is also evident from Eq.\ref{eq:J}, wherein J reduces
to $-2w_{n}\sin(2\theta)/(2w_{n}^{2}+\sin(2\theta)^{2}), E = iw_{n}$.
Fig.~\ref{J-d}(b) shows the current-phase relation for two different
values of the Fermi energy. It again confirms the earlier indication
of $\pi$ shift.  Finally, to establish beyond doubt that as function
of Fermi energy one generates a $\pi$ junction we plot the free energy
in Fig.~\ref{J-d}(c). The plot shows that as one changes the Fermi
energy via a gate voltage one changes the ground state of the junction
from $0$ to $\pi$.

As shown in Fig.~\ref{J-d}(c-d), the Free energy, $F$, has a
minimum at $\phi=\pi$ (for the $\pi$ junction case) and the
variation of F with $\phi$ is strongly dependent on the length $d$
and the Fermi energy. In this parameter regime the Free energy can
be approximated as $F\sim -E_{\pi} [\cos(\phi_{\pi}+\pi)+1]$, with
$E_{\pi}$ being the Josephson coupling constant. The $0$  and
$\pi$ junctions, depicted in Fig. 1, have Josephson energies
$U_{0}=E_{0}|\sin(\phi_0 /2)|$ and
$U_{\pi}=-E_{\pi}[\cos(\phi_\pi+\pi)+1]$ plotted in
Fig.~\ref{J-d}(d). The superconducting phase difference is
$\phi_0$ for the $0$ junction and $\phi_\pi$ for the $\pi$
junction. The total flux in the ring $\Phi$ satisfies
$\phi_{\pi}-\phi_{0}=2\pi\Phi/\Phi_0 -2\pi l$, where $\Phi_0$ is
the flux quantum and $l$ is an integer.
\section{Qubits and Gates}
\begin{figure}
  \centerline{\includegraphics[width=9cm,height=4.5cm]{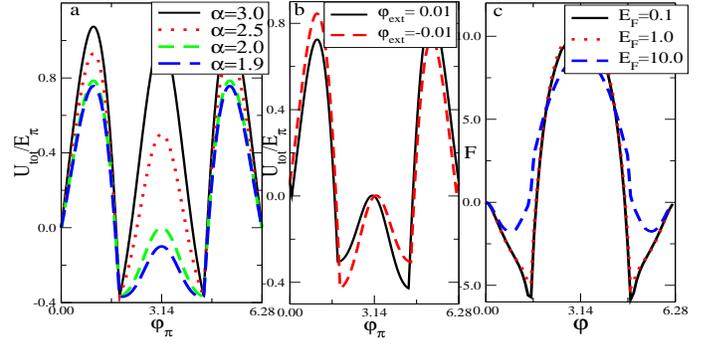}}
  \caption{(Color online) (a) Normalized energy, $U_{tot}/E_\pi$, as function of
    $\phi_\pi$ for no external magnetic flux. (b) In presence of
    an external magnetic field with $\alpha=2.5.$ (c) The
    degenerate ground states of  a d-wave Josephson junction.
    \label{qubit}}
\end{figure}

In Ref.~\onlinecite{maekawa} the authors demonstrate a qubit with a
$\pi$ (SFS) junction\cite{buzdin} and a 0 (SNS) junction coupled into
a ring. In our work we predict that our graphene based system, which
does not need any ferromagnetic element in contrast to
Ref.\onlinecite{maekawa}, could implement a qubit. Further we show how
to implement single qubit gates using our set up. The full Hamiltonian
of the graphene ring system (Fig. 1) is given by $H=K+U_{tot}$ with
$U_{tot}=U_{0}+U_{\pi}+U_L$, where $U_L=(\Phi-\Phi_{ext})^2/2L_S$ is
the magnetic energy stored in the ring and $K$ is the flux independent
kinetic energy. We next minimize the Hamiltonian with respect to flux
and obtain $\Phi(\phi_0)=\beta\Phi_{0}\sin(\phi_\pi)+\Phi_{ext}$, with
$\beta=2\pi E_{\pi} L_S/\Phi_0^2$. Substituting this equation in the
expression for $U_{tot}$, we have:
\begin{equation}
  \frac{U_{tot}}{E_{\pi}}=\alpha[|\sin(\frac{\phi_\pi}{2}
  -\frac{\pi\Phi}{\Phi_{0}})|]+[\cos(\phi_{\pi})-1]+\pi\beta\sin^{2}(\phi_{\pi}).
  \label{minH}
\end{equation}
with $\alpha=E_0 / E_\pi$. For typical values mentioned in
Fig.~\ref{qubit}, we plot Eq.~(\ref{minH}).
We observe that the energy has double minima located approximately
at $\phi_\pi\sim 3\pi/5$ ($|0\rangle$ state) and $7\pi/5$
($|1\rangle$ state) which form the basis of the qubit. For single
layer graphene with junction area~\cite{heersche} $0.8 \times
10^{-12} \mu$ m$^{2}$ and depth $1$ nm, the electrostatic energy
$E_{c}$ is $2.5 \times 10^{-24}J$, while $E_0$ the junction energy
for the zero junction is around $1000 E_c$. Thus for $\alpha=3.0$,
we have $\Delta E$, the energy gap, between the ground and first
excited state $\Delta E/h = 1000 $GHz. The basic phase gate with
$\phi=\Delta E \Delta t/\hbar=\pi$ could be implemented with gate
time $\Delta t$ given by $1$ pico-second. In Fig.~4(c), the Free
energy of a basic d-wave graphene Josephson junction is plotted
for different values of Fermi energy and width $d=0.001$. One can
easily see that degenerate states are formed at $\phi\sim\pi/2$
and $3\pi/2$. The coupling between these states can be easily
varied by the gate voltage effectively realizing single qubit
gates as aforementioned.

\begin{figure}
\vskip 1.0cm
  \centerline{\includegraphics[width=9cm,height=5cm]{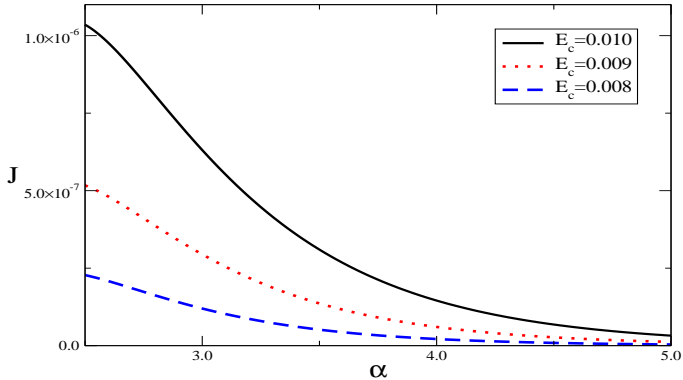}
  } \caption{(Color online) Exchange coupling J (in units of $E_\pi$) as function
    of $\alpha$ for different values of $E_{c}$. Here $E_c \ll E_\pi$
    .\label{exchange}}
\end{figure}

We will now show how to implement an exchange
gate $\sigma_x$ acting on the qubit states $|0\rangle$ and
$|1\rangle$ for the structure as depicted in Fig.~1(a). This is
realized by a tunnelling transition between the potential minima
that encode these qubit states. Assuming the coupling potential is
deep enough we approximate the qubit states by Gaussians centered
at the minima of $U_{tot}$. By varying $\alpha$ (or $E_c$) one can
induce tunnelling between the two minima in a controlled way. The
exchange coupling of our system is calculated as
\begin{equation}
  J=\int d\phi_{\pi}
  \Psi^{*}(\phi_{\pi}-\phi_{|0\rangle})
  \Big(-4E_{c}\frac{d^2}{d\phi_{\pi}^2}+U_{tot}\Big)
  \Psi^{}(\phi_{\pi}-\phi_{|1\rangle}).
\end{equation}
In Fig.~\ref{exchange} we plot the exchange coupling versus the
normalized Josephson energy for various values of the
electrostatic energy, $E_c$ in units of $E_\pi$. We see that for
large $\alpha$ no tunnelling occurs, while for $\alpha\sim 3.0 $
we obtain $J \sim 10^{-6} E_{\pi}$ (for $E_{c}=0.01$) and, thus,
the $\sigma_x$ gate can be implemented in $\Delta t \sim 10^{-6}$
seconds.

To conclude we have shown a novel implementation of a Josephson qubit
using graphene as a substrate. Our work is the first to predict a
qubit using only monolayer graphene. It was shown that a ferromagnetic
graphene layer is unnecessary to create a $\pi$-shift, a completely
novel result. $\pi$ junctions have special role in a host of
applications ranging from their use in superconducting digital
circuits to superconducting qubits. We have shown how a $\pi$ junction
is formed in graphene where it can be very easily tuned by the
application of a gate voltage alone. Secondly, we propose Josephson
qubits and we present the phase and exchange gates for quantum
computation purposes. Future proposals to make CNOT or other two-qubit
gate designs could also be envisaged using the above architecture.

\section{Acknowledgements} The authors acknowledge useful correspondence
  with Carlo Beenakker on a previous version of this manuscript. This
  work was supported by the EU grants EMALI and SCALA, EPSRC and the
  Royal Society.

\end{document}